**Translational-Rotational Coupling during the Scattering of a Frictional Sphere from a Flat Surface**

Yueran Wang and Peter Harrowell[*]

School of Chemistry, University of Sydney, Sydney 2006, New South Wales, Australia

* Corresponding author: peter.harrowell@sydney.edu.au

Abstract

At a macroscopic level, concepts such as 'top spin', 'back spin' and 'rolling' are commonly used to describe the collision of balls and surfaces. Each term refers to an aspect of the coupling of rotational motion during the collision of a spherical particle with a planar surface. In this paper we explore the mechanisms of energy transfer involving the collision of a rotating sphere and a surface using a model of frictional interactions developed for granular material. We present explicit analytical treatments for the scattering and derive expressions for two important limiting classes: energy conserving collisions and collisions subject to rapid transverse dissipation.

**1. Introduction**

The collisional coupling involving particle rotations occurs because the anisotropic component of the particle interaction potential couples the force on the centre of mass with a torque. Our goal in this paper is to develop an explicit physical account of how this coupling takes place. What we mean by 'how', in this case, is encapsulated in the explicit relationship between the velocities (translational and rotational) after the collision in terms of the initial values. Such relations are called 'kinematic' in that they do not include the detailed dynamic trajectory of the collisional event. Obtaining these kinematic descriptions of collisions is made difficult by the very anisotropies on which the translational-rotational (TR) coupling relies. The outcome of a collision (for simplicity we shall consider the particle-surface collision) will result in a distribution of final velocities rather than a single set of values, even when the initial velocities are precisely determined, due the distribution of initial orientations of the particle. Examples of this distribution for the classical scattering of a diatomic from surface are provided in refs. [1-5]. As the anisotropy of the particle increases, so too will the width of the distribution of final states [6]. A second difficulty with particle anisotropy is the unbounded diversity of particle shapes and the possibility of an equally diverse range of possible kinematic outcomes. How should one choose what is a general feature of the TR coupling and what is specific to the details of the interaction?

To address these difficulties of particle anisotropy, we have elected to study the TR coupling of a spherical particle with a frictional interaction with the scattering surface. Friction provides a means of producing a torque on the particle, analogous to that resulting from anisotropies, while avoiding the attendant complications of distinguishable particle orientations. As described in Section 2, we shall make use of a model, developed for granular materials by Cundall and Strack [7], that includes a frictional contact similar to that observed between macroscopic particles. We make no claim that this simple model will describe the details of any particular molecular collision but, hopefully, it will capture some aspects common to all such collisions.

There is a considerable literature on collisions with friction, a literature distributed across a diverse range of fields including sports physics [8-10], mechanical engineering [11-12], applied mathematics [13] and robotics [14]. In 1687, Newton [15] proposed a kinematic equation for inelastic collisions, $v_n(\infty) = -\varepsilon v_n(0)$, where $v_n(\infty)$ and $v_n(0)$ are the final and initial normal components of the translational velocity and $\varepsilon$ is the empirical coefficient of restitution with a value between 0 and 1. The dissipation reflected in $\varepsilon < 1$ is associated with the deformation of the colliding object. In the case of oblique (i.e. non-collinear) collisions, there is a non-zero transverse component of the relative translational velocity that results in sliding during the collision. If the surfaces are 'rough', this sliding results in a transverse force due to friction which, in turn, generates a torque that couples to the angular velocity of the particle [16]. There is a certain arbitrariness as to whether this roughness is to be treated explicitly in terms of shape anisotropy or, implicitly, in the form of a surface friction. As demonstrated by Zwanzig [17] in the context of particles in liquids, the rotation of an anisotropic cylinder with a smooth surface (i.e. slip boundary conditions) behaves very similarly to that of a circular cylinder with a rough surface (i.e. stick boundary conditions). This qualitative equivalence provides a loose rationale for our strategy in this paper of studying frictional spheres in place of anisotropic particles.

The goal of this paper is to derive kinematic expressions for the collision of a rotating frictional particle with a flat surface that will explicitly describe the translational-rotational coupling. In doing so we shall derive two interesting limits: the purely elastic collision of a rough sphere and the rolling contact in the limit of rapid inherent dissipation, and explicitly account for the different channels of energy dissipation arising from the frictional collisions.

The paper is organised as follows. In Section 2 we describe the equations of motion for the frictional particle. In Section 3 we present the analytic solutions for the periodic evolution of the transverse and angular velocities under the constraint of fixed normal distance from the surface, the 'frozen' collision approximation. In Section 4 we use the results of the previous section to derive the general kinematic description of the particle-surface collision and establish the conditions under which 'rough' sphere dynamics is recovered. In Section 5 we consider the energy dissipated during the collision, derive the kinematic expressions for damped collisions and the conditions under which rolling collisions occur.

## 2. Frictional Model and Granular Dynamics

Treatments of the dynamics of spherical particles interacting via surface friction have been developed for the study of granular material. The Cundall-Strack model [7] has been widely used to model the effect of friction in a variety of granular applications [18]. In 2001, Silbert et al [19] presented an implementation of this model, available in the LAMMPS package [20], that we shall use in this paper.

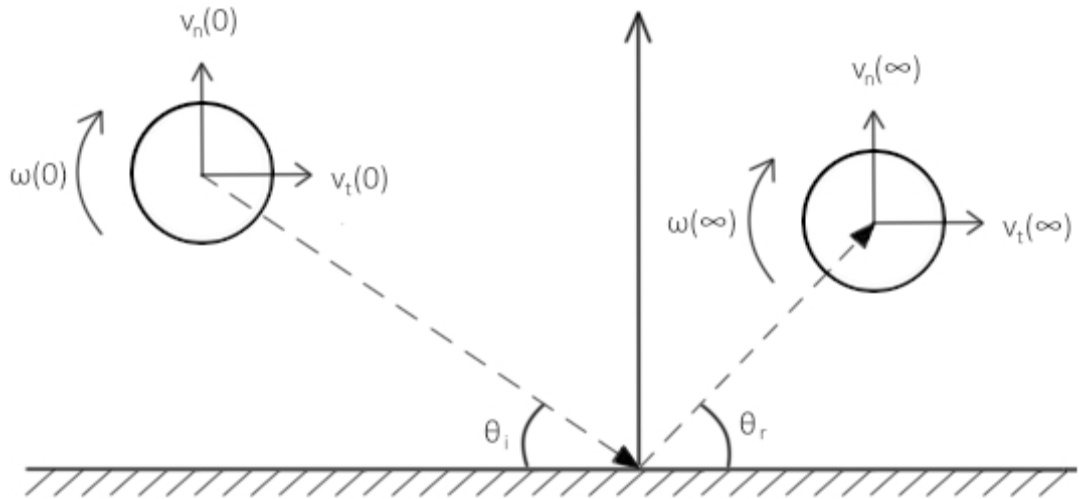

**Figure 1.** A sketch of the scattering geometry indicating the initial and final components of the normal and transverse velocity, $v_n(0)$, $v_n(\infty)$, $v_t(0)$ and $v_t(\infty)$, respectively and the initial and final angular velocities, $\omega(0)$ and $\omega(\infty)$. Note that angles of incidence and reflection are given by $\theta_i = \tan^{-1}\left(\dfrac{-v_n(0)}{v_t(0)}\right)$ and $\theta_r = \tan^{-1}\left(\dfrac{v_n(\infty)}{v_t(\infty)}\right)$, respectively.

The different degrees of freedom of the particle-surface scattering are sketched in Fig. 1. The translational velocity $\vec{v}$ associated with the motion of the centre of mass of the sphere can be resolved into normal and transverse components, $v_n$ and $v_t$, respectively, as show. The sign conventions in vector algebra (the 'right hand rule') specifies that a positive angular velocity corresponds to an anti-clockwise rotation in the plane shown in Fig. 1 and a negative value corresponds to a clockwise motion.

The granular model of ref. [7,19] is defined by the description of the forces between particles (or, in this case, particle and surface). When resolved into their normal and transverse components, the particle-surface forces are defined as follows:

$$F_t = h(1 - z/r)[-k_t u - \gamma_t \dot{u}]$$
$$F_n = h(1 - z/r)[-k_n(z-r) - \gamma_n \dot{z}] \qquad (1)$$

where $k_t$ and $k_n$ are the transverse and normal force constants, $\gamma_t$ and $\gamma_n$ are explicit friction coefficients, u is the elastic slip displacement between the point of contact between sphere and surface at time t as defined below, z is the normal displacement measured from the surface position at z = 0 (i.e. $\dot{z} = v_n$) and r is the radius of the particle. The function $h$ determines the spatial extent of the particle-surface interaction and is given here by the linear spring condition, $h(x) = 1$ for $x \geq 0$ and 0 otherwise. In the following we shall set the friction coefficients $\gamma_t$ and $\gamma_n$ to zero except, as indicated, in Section 5. The mass of the particle is $m$.

The rate of change of the surface displacement u can be written as

$$\dot{u} = v_t + r\omega \qquad (2)$$

where $v_t$ is the transverse component of the particle velocity. The quantity $\dot{u}$ is the slip between the particle and the surface. The condition for rolling contact is zero slip, i.e.

$v_t + r\omega = 0$. This transverse force acts directly on the transverse and angular velocities via the equations,

$$I\dot{\omega} = rF_t \tag{3}$$

$$m\dot{v}_t = F_t \tag{4}$$

where I is the moment of inertia of the particle. Eq.3 is a straightforward reduction of the general expression for the action of a torque expressed as the tangential force. The one subtlety being the choice of the radial length *r*. The value of *r* in Eq. 3 corresponds to the distance between the point of application of force $F_t$ and the centre of the sphere. This distance might be different from the radius of the sphere used to determine the moment of inertia *I*. It is the second equation where the granular model [7,19] seriously modifies the physics of the problem. If the tangential force acts on the surface of the sphere at the point of contact, it cannot, strictly, be acting directly on the center of the sphere as well. The kinematic linkage between the point of contact and the particle center has been considered by a number of groups [11, 20]. Unless otherwise stated, we use the following parameter values: *m = 1, r = 0.5, $k_n$ = 10 and $k_t$ = 1*.

## 3. Solutions for the 'Frozen' Collision Approximation

The two components of the translational velocity evolve quite differently during the collision. The transverse velocity $v_t$ couples to the rotation of the sphere in the plane containing $\vec{v}_t$ and the surface normal $\hat{\vec{n}}$. The normal velocity, $v_n$, describes the approach and recoil from the surface and hence determines the duration of the collision. The evolution of $v_n$, uncoupled from the rotation, is elastic within the contact distance set by *h(x)* in Eq. 1 . The distinct roles of $v_t$ and $v_n$ in the collision suggest that we can usefully treat them separately. In this Section we shall consider the case where we fix the sphere to sit at a fixed normal distance from the surface, effectively constraining $v_n = 0$. This allows us to describe the collective motion associated with the coupling between $v_t$ and the rotational velocity, $\omega$. We shall refer to this constrained situation as the frozen collision assumption.

A direct consequence of Eqs. 3 and 4 is the following relation between the two degrees of freedom,

$$\frac{I}{r}\dot{\omega} = m\dot{v}_t \tag{5}$$

On integration of Eq. 5, we find the following relation between the two degree of freedom

$$\Delta v_t(t) = \frac{I}{mr^2} r\Delta\omega(t) \tag{6}$$

where $\Delta v_t(t) = v_t(t) - v_t(0)$ and $\Delta\omega(t) = \omega(t) - \omega(0)$. There are limited options for the moment of inertia for a spherical particle. In this model, the moment of inertia represents the only aspect of the particle property that can be adjusted. If we write the moment of inertia $I = \alpha mr^2$ then we could imagine α varying from zero (all the mass located at the point center) to 2/3 (where the mass is localised in an infinitesimal shell at the surface). In principal, larger values of α could be used but these would require us to allow for 'phantom' mass that lay beyond the radius of the sphere. If we treat the sphere as a uniform solid then $\alpha = \frac{2}{5}$.

To solve for the explicit time dependence of the transverse and angular velocities we must consider the evolution of the displacement u. Combining Eqs. 1-4 we have

$$\ddot{u} = \dot{v}_t + r\dot{\omega} = -k_t\left[\frac{1}{m} + \frac{r^2}{I}\right]u = -Qu \tag{7}$$

where $Q = k_t \left[ \dfrac{1}{m} + \dfrac{r^2}{I} \right] = \dfrac{k_t}{m} \dfrac{1+\alpha}{\alpha}$ (8)

with a solution

$$u(t) = \dfrac{\dot{u}(0)}{\sqrt{Q}} \sin\left(t\sqrt{Q}\right)$$ (9)

where $\dot{u}(0) = v_t(0) + r\omega(0)$. Substituting Eq. 9 into Eqs. 3 and 4, we can now solve for $v_t(t)$, i.e.

$$\begin{aligned} v_t(t) &= v_t(0) + \dfrac{k_t \dot{u}(0)}{mQ}\left[\cos\left(t\sqrt{Q}\right) - 1\right] \\ r\omega(t) &= r\omega(0) + \dfrac{r^2 k_t \dot{u}(0)}{IQ}\left[\cos\left(t\sqrt{Q}\right) - 1\right] \end{aligned}$$ (10)

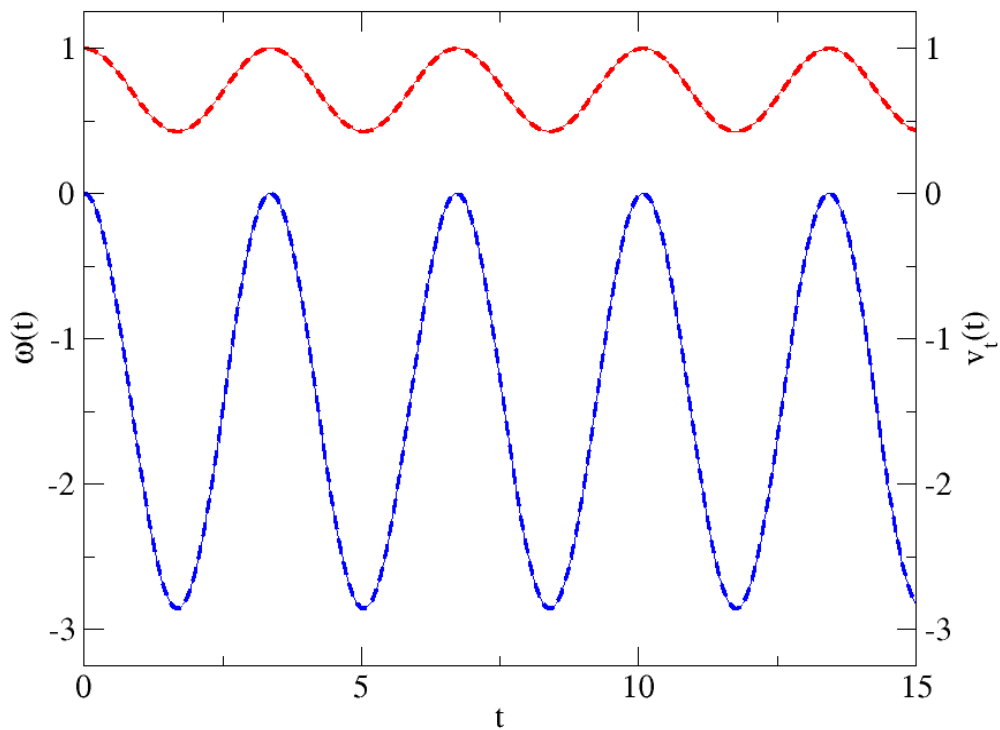

**Figure 2.** The time evolution of $v_t$ (red) and $\omega$ (blue) for a particle constrained to a vertical distance of $z = 0.1$ from the surface. The initial conditions are $v_t(0) = 1.0$ and $\omega(0) = 0$. Two

curves are shown, one from the numerical integration (dashed line) and the second from Eqs. 10 (solid line).

By fixing the normal distance of the particle from the surface, we have effectively 'frozen' this degree of freedom. The simple oscillations described by Eq. 10 and Fig. 2 provide an explicit representation of the translational-rotational coupling in this frozen collision. Essentially, both modes have lost their original identities to become, by virtue of the influence of the transverse force, coupled vibrations. The transverse velocity is periodically diminished as the magnitude of the rotational velocity increases. According to Eq. 10, $v_t$ and $r\omega$ oscillate about non-zero mean values. From Eq. 10, these mean values are

$$\bar{v}_t = v_t(0) - \dot{u}(0)\frac{\alpha}{1+\alpha}$$
$$r\bar{\omega} = r\omega(0) - \dot{u}(0)\frac{1}{1+\alpha} \qquad (11)$$

From Eq.11 we have $\bar{v}_t + r\bar{\omega} = 0$, in other words the mean values correspond to the rolling solution. This result underscores the point that the transverse force exerted by the surface should not be thought of as a spring attached to a point on the particle's surface. The motion described by Eq.10 and Fig. 2 can be best summarised as rolling accompanied by an oscillating slip. It is the rolling motion that gives rise to the persistent transverse translation of the particle, while the slip, u(t), drives the transfer of energy between the particle rotation and its transverse motion.

The coupling of the translational and rotational velocities in the frozen collision can be explicitly illustrated by the trajectory in the phase space [$v_t$,r$\omega$]. Examples of the family of phase space curves for the case where r$\omega$(0) is fixed at -0.5 are plotted in Fig. 3 for a range of values of $v_t(0)$. We find that, in each case, the motion takes place on a line in the phase space between the two extremes of motion, namely $(v_t(0), r\omega(0))$ and $(v_t^m, r\omega^m)$, where

$$v_t^m = v_t(0) - 2\frac{k_t \dot{u}(0)}{mQ}$$

$$r\omega^m = r\omega(0) - 2\frac{r^2 k_t \dot{u}(0)}{IQ}$$
(12)

The fact that each trajectory lies on a straight line is a direct consequence of the assumption, already discussed, of setting the transverse force on the centre of mass equal to that experienced at the particle surface, effectively forcing the translation and rotation to remain in phase. Replacing this simplification by a more realistic coupling (see ref. [11,20]) would see each straight line in Fig. 3 replaced by an elliptical orbit. From Eq. 6, the slope of the lines in Fig. 3 are given by

$$\frac{dv_t}{d(r\omega)} = \frac{I}{mr^2} = \alpha$$
(13)

which, for the case here of the solid sphere equals 2/5, in agreement with the observed slope in Fig. 3. To summarise, the coupling between the transverse translation and the rotation, as measured by the slope (Eq. 12) depends only on the distribution of the mass m in the particles, as characterised by the moment of inertia I. The details of the interaction, e.g. the value of $k_t$, establishes the bounds on the values of $v_t$ and $\omega$.

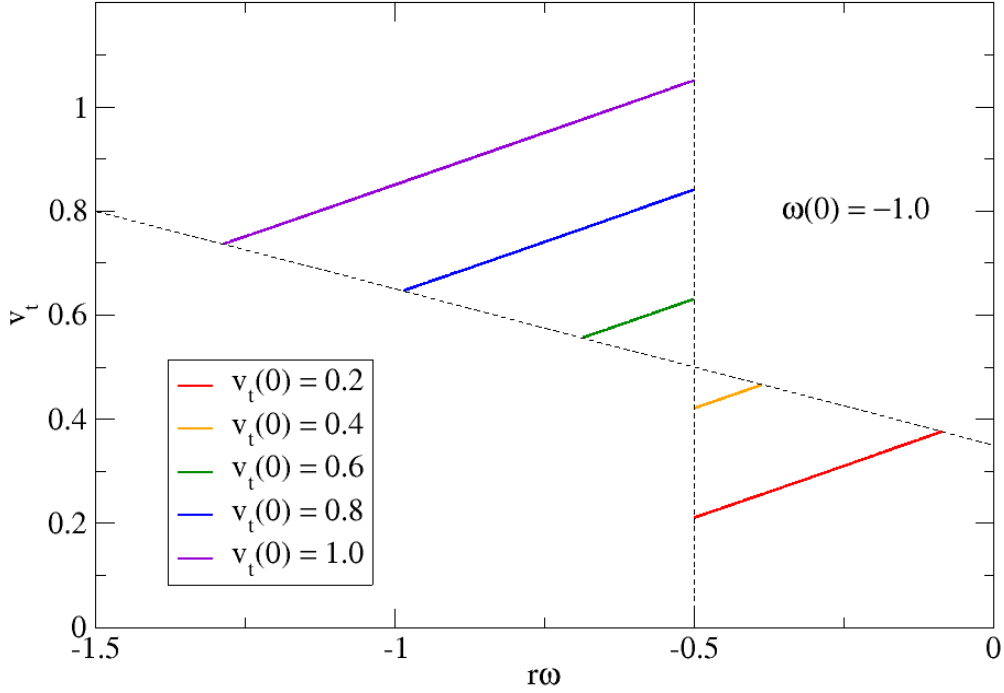

**Figure 3.** The trajectories of particles in the space of $v_t$ vs $r\omega$ held at a fixed normal position of $z = 0.1$ for different values of the initial transverse velocity $v_t(0)$. In each case, $r\omega(0) = -0.5$.

## 4. Finite Interaction Times and the Scattering Matrix

Having established what can happen in terms of TR coupling while the particle and wall are in contact, we shall consider in the Section how this information helps us understand the outcome of the overall collision event. Specially, if the quantities $v_t$ and $\omega$ can takes on any set of values along a line like those shown in Fig. 3, which set of values ends up as the outcome of a collision event?

On restoring the non-zero velocity in the normal direction, the particle will only be in the vicinity of the interface from some finite collision time τ. The overall collision can, then, be regarded roughly as the vibrations depicted in Figs. 2 and 3, interrupted after a time τ. The duration of a collision is determined by the recoil of particle from the surface along the normal direction. In general, this recoil will remove the particle from the wall and the associated transverse force at a point other than the extremes of the motion. The normal velocity obeys the following Hookean equation of motion,

$$m\dot{v}_n = k_n(r-z) \text{ for } z \leq r \tag{14}$$

and zero for $z > r$ where z is the normal distance between the particle and the surface and $z > 0$ corresponds the free space above the surface. Note that the transverse interaction, as expressed in Eq. 1, also vanishes when $z > r$. The collision time $\tau$, therefore, is the time between the two instances of z = r. We can solve Eq.14 to give

$$z(t) = r + \frac{v_n(o)}{p}\sin(pt) \tag{15}$$

where $p = \sqrt{\frac{k_n}{m}}$. It follows that time $\tau$ required for a particle to return to its initial distance from the surface is

$$\tau = \frac{\pi}{p} = \pi\sqrt{\frac{m}{k_n}} \tag{16}$$

Note that this time is independent of the initial velocity $v_n(0)$. This independence is a particular consequence of the harmonic interaction and, in general, the collision time will be a decreasing function of $v_n(0)$. In the case of a hard sphere-wall interaction, for example, we would have $\tau = 2r/v_n(0)$. Substituting $\tau$ from Eq. 15 into Eq. 10 and equating $v_t(\infty) = v_t(\tau)$ and $r\omega(\infty) = \omega(\tau)$, we can now estimate the actual final velocities and the associated TR coupling for an arbitrary collision, as

$$v_t(\infty) = v_t(0) + \frac{k_t \dot{u}(0)}{mQ}\left[\cos\left(\pi\sqrt{\frac{mQ}{k_n}}\right) - 1\right]$$

$$r\omega(\infty) = r\omega(0) + \frac{r^2 k_t \dot{u}(0)}{IQ}\left[\cos\left(\pi\sqrt{\frac{mQ}{k_n}}\right) - 1\right] \quad (17)$$

Inserting $\dot{u}(0) = v_t(0) + r\omega(0)$, we can rewrite Eq. 17 in matrix form as

$$\begin{pmatrix} v_n(\infty) \\ v_t(\infty) \\ r\omega(\infty) \end{pmatrix} = \begin{pmatrix} -1 & 0 & 0 \\ 0 & 1+\frac{\alpha\Omega}{1+\alpha} & \frac{\alpha\Omega}{1+\alpha} \\ 0 & \frac{\Omega}{1+\alpha} & 1+\frac{\Omega}{1+\alpha} \end{pmatrix} \begin{pmatrix} v_n(0) \\ v_t(0) \\ r\omega(0) \end{pmatrix} \quad (18)$$

where $\Omega = \cos\left(\pi\sqrt{\frac{k_t}{k_n}\frac{1+\alpha}{\alpha}}\right) - 1$. The off-diagonal terms in the scattering matrix in Eq. 18 represent explicit measures of the translation-rotation coupling in this model.

The relationship between $v_t$ and $\omega$ in Eq. 14 can be recast as follows:

$$\Delta v_t = v_t(\infty) - v_t(0) = \frac{\alpha\Omega}{1+\alpha}\dot{u}(0)$$

$$r\Delta\omega = r\omega(\infty) - r\omega(0) = \frac{\Omega}{1+\alpha}\dot{u}(0) \quad (19)$$

to clarify that the coupling between the transverse velocity and the particle rotation is directly proportional to the initial slip velocity $\dot{u}$. Note that $\Omega \leq 0$ for all values of $k_t$ and $k_n$. Here we have used $k_t = 1$ and $k_n = 10$, and, hence, we have $\Omega = -1.2838$ (when $\alpha = 2/5$). As Eq. 19 makes clear, there is no overall change in $v_t$ or $\omega$ when $\dot{u} = 0$, i.e. the rolling condition still ensures that no TR coupling occurs even for collisions of arbitrary duration.

To test the accuracy of Eq. 18 and 19, we have, in Fig. 4, plotted $\Delta v_t$ and $r\Delta\omega$ against $\dot{u} = v_t(0) + r\omega(0)$ calculated by trajectory calculations for a range of initial conditions and compared the results against the straight lines predicted in Eq. 19. We find excellent agreement between the numerical integration of the dynamic equations and the derived kinematic scattering matrix in Eq. 18.

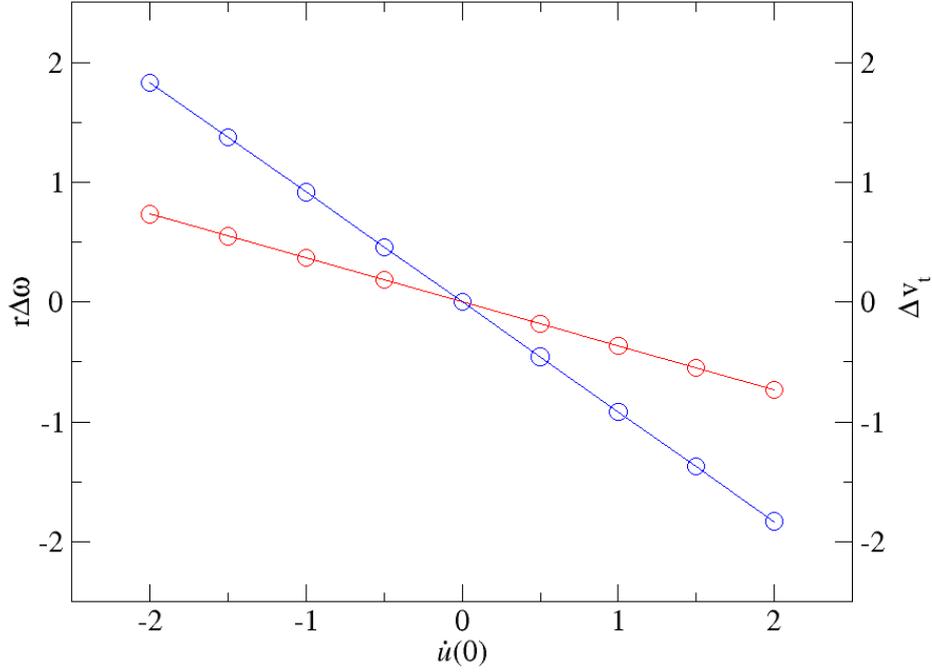

**Figure 4.** Scatter plots of $\Delta v_t$ (blue circles) and $r\Delta\omega$ (red circles) vs $\dot{u} = v_t(0) + r\omega(0)$ for a variety of different choices of initial conditions. Also plotted are the straight lines predicted in Eq. 19 for $\Delta v_t$ (blue line) and $r\Delta\omega$ (red line).

In the Cundall-Strack [7] equations of motion, the transverse force responsible coupling the translational and rotational motion arises in response to the surface slip displacement u(t). In general, when the contact between sphere and surface is broken by the recoil of the particle during the collision, this displacement will be some non-zero amplitude. This means that the potential energy associated with the slip will be lost as dissipated energy. We shall return to this lost energy in the next Section. Here we consider the possibilities of energy conserving scattering events. To avoid losing energy the slip displacement must be zero at the point at which the particle loses contact with the wall. From Eq. 9 we find that these zeros in u(t) occur at $t = n\frac{\pi}{\sqrt{Q}}$ for n = 0,1, …. The case of n = 0, 2, 4 … simply correspond to $v_t(\infty) = v_t(0)$ and $\omega(\infty) = \omega(0)$, i.e. the slip wall with no tangential force acting. A more interesting result is found when we let n be odd. To achieve a collision time $\tau = \frac{\pi}{\sqrt{Q}}$ (i.e. n=1) we must set $\frac{k_t}{k_n} = \frac{\alpha}{1+\alpha}$. With this choice, the quantity $\Omega = -2$ and the scattering matrix becomes

$$\begin{pmatrix} v_n(\infty) \\ v_t(\infty) \\ r\omega(\infty) \end{pmatrix} = \begin{pmatrix} -1 & 0 & 0 \\ 0 & 1-\frac{2\alpha}{1+\alpha} & -\frac{2\alpha}{1+\alpha} \\ 0 & -\frac{2}{1+\alpha} & 1-\frac{2}{1+\alpha} \end{pmatrix} \begin{pmatrix} v_n(0) \\ v_t(0) \\ r\omega(0) \end{pmatrix} \quad (20)$$

Eq. 20 described the kinematics of bouncing of a 'rough' sphere from a surface. This expression was derived by Garwin [22] in 1969 based entirely on the conservation of energy and momentum. This rough sphere model has been considered at some length as a model of TR coupling in liquids. Chapman and Cowling [23] derived the kinematic expressions for the collisions between two rough spheres. In a series of papers [24-26], Berne and coworkers studied the velocity correlations in molecular dynamics simulations of rough sphere liquids. Note that Eq. 20 contains no information about the details of the transverse force between particle and surface, only the parameter α characterising the mass distribution with the sphere. In the case of a rolling contact, i.e. $v_t(0) + r\omega(0) = 0$, Eq. 20 reduces to frictionless surface result, an observation consistent with the fact that the rolling particle experiences no transverse force.

The behavior of the rough sphere is quite different from that of a sphere bouncing from a slipping surface (i.e. one with no translation-rotation coupling). In Fig. 5, we sketch some example trajectories of the rough sphere scattering. Consider the case of solid particle with zero initial rotation, $\omega(0) = 0$. According to Eq. 20, the rotational velocity after the collision is $r\omega(\infty) = -\frac{10}{7}v_t(0)$ while the final value of $v_t$ is reduced. As already noted, a value $\omega < 0$ corresponds to a clockwise rotation which, in this case, resembles a forward roll. The transfer of momentum out of the transverse translational motion results in an angle of reflection smaller than the angular of incidence. Similarly, a spinning solid particle undergoing a normal impact, i.e. $v_t(0) = 0$, will bounce off with a transverse velocity of $v_t(\infty) = -\frac{1}{7}r\omega(0)$ as indicated in Fig. 5. A close realization of the rough sphere is the toy sold as the *Wham-O Super-ball*® and a video [27] of superball trajectories provides a clear illustration of the characteristic bounces depicted in Fig. 5.

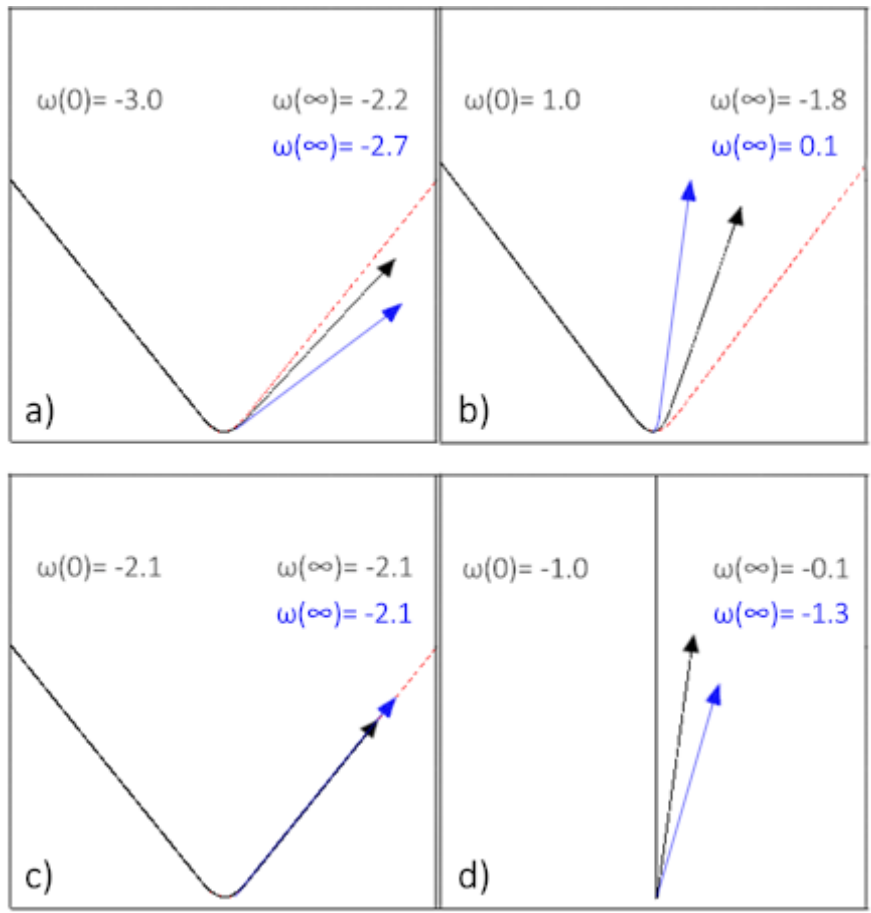

**Figure 5.** Examples of scattering trajectories governed by Eq. 20 (rough sphere – black lines) and by Eq.23 ('rolling' collision - blue lines). The initial rotational velocity corresponds to a) forward (clockwise) rotation (i.e. 'top spin'), b) backward rotation (i.e. 'backspin'), and c) a value corresponding to the rolling condition (see text). In d) the initial transverse velocity $v_t(0) = 0$ but the final value does not due to transfer of momentum from the initial rotation to translation. The trajectories in the absence of transverse force are indicated as dashed lines (red).

## 5. Energy Dissipation

Energy dissipation arises from two features of the equations of motion in Eq. 1. The first is explicit in the form of a friction coefficients $\gamma_n$ and $\gamma_t$ in Eq. 1. The second source of dissipation is the energy lost when a collision ends with a non-zero elastic strain associated with an unrelaxed elastic slip displacement. The energy loss ΔE during a collision can be calculated as the difference in final and initial kinetic energies, i.e.

$$\Delta E = \frac{m}{2}\left(v_t^2(\infty) - v_t^2(0)\right) + \frac{I}{2}\left(\omega^2(\infty) - \omega^2(0)\right) + \frac{m}{2}\left(v_n^2(\infty) - v_n^2(0)\right) \quad (21)$$

We can define the dissipation as the fractional $f$ amount of the initial kinetic energy lost in the collision, $f = -\dfrac{\Delta E}{E(0)}$. Since any energy loss is associated with the involvement of surface slip, it is useful to consider the relationship between $f$ and the initial transverse slip rate, i.e. $\dot{u}(0) = v_t(0) + r\omega(0)$ as presented in Fig. 6. We note that $f = 0$ when $\dot{u}(0) = 0$. Data based on two different values of the ratio of force constants are presented, $k_t/k_n = 1/10$ and $2/7$, where the latter value corresponds to the rough sphere behaviour described in Eq. 20. The absence of dissipation for $k_t/k_n = 2/7$, while following directly from the energy conservation condition used in its derivation, is remarkable in that it is achieved purely through adjusting the relative strengths of the transverse and normal interactions.

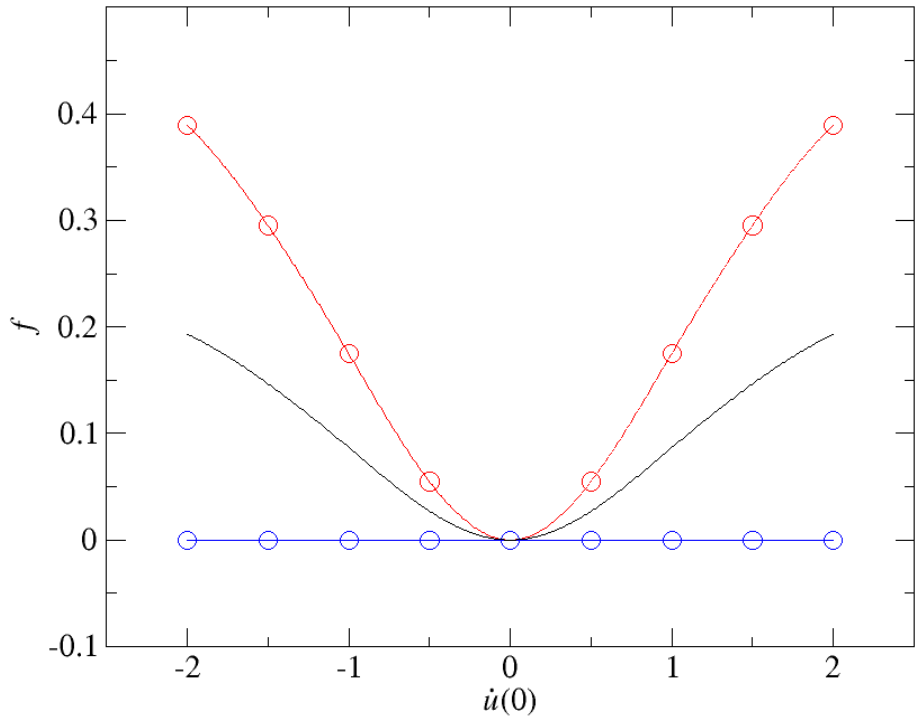

**Figure 6.** A plot of the dissipation $f = -\frac{\Delta E}{E(0)}$ as a function $\dot{u}(0)$ for a range of initial values of $\omega$, $v_t$ and $v_n$. The results from the simulation are presented as circles and the solid curve is the energy loss predicted by Eq. 18. Results are presented for $k_t/k_n = 1/10$ (red) and $2/7$ (blue), along with the results from Eq. 23 for the rolling collision corresponding to the limit of a large transverse damping coefficient $\gamma_t$ (black curve).

In this model we have ignored the phonon modes of the surface. Once consequence of these modes would be to provide an irreversible sink for energy from the colliding particle. The granular equations in Eq. 1 allow this effect to be modelled through the explicit frictions $\gamma_t$ and $\gamma_n$. We shall consider the consequence of a non-zero value of the transverse friction $\gamma_t$. (The effect of a non-zero $\gamma_n$ would be to increase the collision time $\tau$ and make this time dependent on $v_n(0)$ but without introducing any qualitative change in the TR coupling.) In Fig. 7 we plot the time dependence of $v_t$ and $r\omega$ in the frozen collision limit (i.e. calculations analogous to those presented in Fig. 2) for a range of values of $\gamma_t$. In all cases the velocities decay to their mean value, provided in Eq. 11, corresponding to the rolling particle. The decay can be expressed by multiplying the $\cos(t\sqrt{Q})$ in Eq. 10 by $\exp(-\gamma_t t/m)$ and, hence, replacing the factor $\Omega$ in Eq. 18 by

$$\tilde{\Omega} = \exp\left(-\frac{\pi\gamma_t}{\sqrt{mk_n}}\right)\cos\left(\pi\sqrt{\frac{k_t}{k_n}\frac{1+\alpha}{\alpha}}\right) - 1 \qquad (22)$$

It follows that for $\gamma_t/m \gg \tau^{-1}$, where $\tau$ is the collision time given by Eq. 16, $\tilde{\Omega} = -1$ and the outcome of the scattering corresponds to the rolling condition as determined by Eq. 11, i.e.

$$\begin{pmatrix} v_n(\infty) \\ v_t(\infty) \\ r\omega(\infty) \end{pmatrix} = \begin{pmatrix} -1 & 0 & 0 \\ 0 & 1-\dfrac{\alpha}{1+\alpha} & -\dfrac{\alpha}{1+\alpha} \\ 0 & -\dfrac{\alpha}{1+\alpha} & 1-\dfrac{\alpha}{1+\alpha} \end{pmatrix} \begin{pmatrix} v_n(0) \\ v_t(0) \\ r\omega(0) \end{pmatrix} \qquad (23)$$

The possibility that particle-surface collision could result in the particles recoiling with the rolling condition satisfied was the basis of a kinematic treatment proposed by Brody [28] in 1984. Examples of the rolling collision trajectories are provided in Fig. 5 where we find that the rolling collisions exhibit even large deviations from the frictionless case than do the energy conserving collisions of the rough sphere. In Fig. 6 we plot the dissipation $f$ for the rolling collisions and observe that the inclusion of explicit dissipation (i.e. $\gamma_t > 0$) can actually reduce the total dissipation $f$ by reducing the energy lost when recoil occurs with elastic energy still present in the transverse slip.

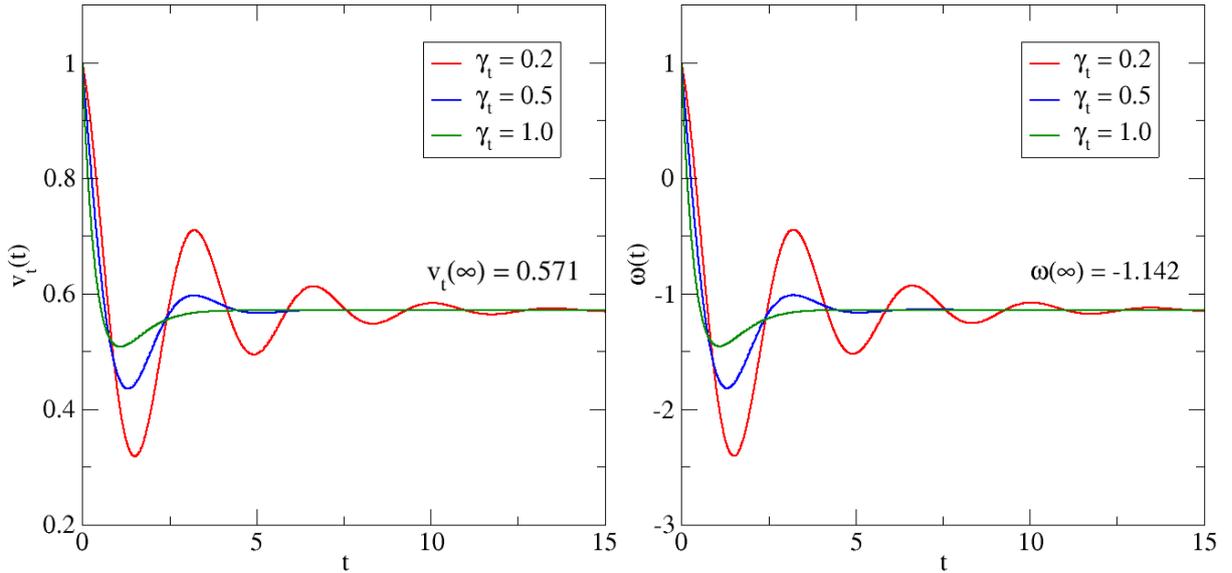

**Figure 7.** The time evolution of a) $v_t$ and b) $\omega$ for a particle constrained to a vertical distance of z = 0.1 with initial conditions $v_t(0) = 1.0$ and $\omega(0) = 1.0$ for three values of the transverse friction as indicated. Note the convergence of the trajectories to the rolling condition.

## 6. Conclusions

The goal of this paper was to present a physical picture of the coupling between translational and rotational motion resulting from a collision involving a frictional sphere. The physical picture we arrive at is that, for the duration of the collision, translations and rotations lose their very different characteristics and are converted to set of coupled vibrations about the rolling solution. Translation-rotation coupling, therefore, refers to the exchange of momenta between transiently coupled vibrational degrees of freedom. As the normal velocity is nonzero, the duration of the collision is finite and the final state of the scattered particle depends on where along the coupled vibration trajectory the particle-surface system was when the collision ended. In general, the collision ends with elastic energy still stored in the slip displacement, energy that is thus dissipated. This energy loss can be avoided by adjusting the interaction parameter such that the collision time equals half the period of the translation-rotation oscillation. This leads to the surprising result that the frictional particle, as modelled by ref. [7], can undergo collisions that conserve energy and momenta, i.e. the rough particle model, just by selecting the 'magic' ratio of force constants, $\frac{k_t}{k_n} = \frac{2}{7}$ in the case of a solid sphere. This is, we believe, the first example of a spherical model of translational-rotational energy transfer that can be simulated using molecular dynamics with energy conservation as opposed to an event-driven algorithm with the collision outcomes determined by kinematic rules [24-26]. This elastic collision condition appears to be a specific result of the Hookean interaction between particle and wall. Generalising this to a nonlinear interaction will result in a collision time dependent on $v_n(0)$ and, hence, an elastic collision behavior restricted to special values of the initial normal velocity.

With the inclusion of intrinsic dissipation in the equations of motion, we found two striking results. The first was that in the limit of rapid dissipation, the velocities converge to a collisional outcome corresponding to a rolling solution from which no further dissipation is possible. The second result is that, as shown in Fig. 6, the presence of a very large friction during the collision can, for a range of choices of $k_t/k_n$, result in *less* dissipation than the zero friction case. This latter result reflects the two distinct mechanisms for energy loss in the frictional model sed here.

We find the set of rolling solutions appears naturally as a stationary point for the frictional sphere collisions off a single smooth surface. The special status of the rolling solution is erased, however, once we consider particles colliding between multiple surfaces. In the case of parallel surfaces, for example, the only state that can satisfy the rolling condition for both surfaces is the one with zero transverse and angular velocities. Rolling solutions, we note, are a specific feature of particles with circular cross sections.

In this paper we have presented a detailed account of translation-rotational coupling that can, we propose, provide a useful guide for more complex situations. In future work we shall examine how TR coupling influences transport through pores and the nature of TR coupling in liquids of frictional spheres with atomistic interaction potentials.

**Acknowledgements**

The authors gratefully acknowledge support from the Australian Research Council.

**Data Availability**

The data that support the findings of this study are available from the corresponding author upon reasonable request.

**References**

1. J. D. Doll, Simple classical mol of the scattering of a diatomic molecules from a solid surface. *J. Chem.Phys.* **59**, 1038-1042 (1973).

2. I. T. Elsum and R. G. Gordon, A kinematic, classical mechanical theory of reactive collisions. J. Chem. *Phys*. **76**, 3009-3018 (1982)

3. R.B Gerber, Molecular scattering from surfaces: Theoretical methods and results. *Chem. Rev.* **87**, 29-79 (1987).

4. A. J. McCaffery, A new approach to molecular collision dynamics. *Phys. Chem. Chem. Phys.* **6**, 1637-1657 (2004).

5. M. Bonfanti and R. Martinazzo, Classical and quantum dynamics at surfaces: Basic concepts from simple models. *Int. J. Quant. Chem*. **116**, 1575-1602 (2016).

6. W. L. Nichols and J. H. Weare, Rotational energy distributions for homonuclear diatomic beams scattered from solid surfaces: A hard-cube model. *J. Chem. Phys.* **66**, 1075-1078 (1977).

7. P. A. Cundall and O. D. L. Strack, A discrete numerical model of granular assemblies. *Géotechnique* **29**, 47-65 (1979).

8. R. Cross, The bounce of a ball. *Amer. J. Phys.* **67**, 222 (1999)

9. R. Cross, Tennis physics, anyone? *Phys. Today* **61**, 84 (2008).

10. G. Robinson and I. Robinson, The motion of an arbitrarily rotating spherical projectile and its application to ball games. *Physica Scripta* **88**, 018101 (2013).

11. N. Maw, J. R. Barber and J. N. Fawcett, The oblique impact of elastic spheres. *Wear* **38** 101-114 (1976).

12. W. J. Stronge, Rigid body collisions with friction. *Proc. R. Soc. Lond. A* **431**, 169-181 (1990).

13. D. E. Stewart. Rigid-body dynamics with friction and impact. *SIAM Rev.* **42**, 3-39 (2000).

14. S. D. Lee and J. Song, Sensorless collision detection based on friction model for a robot manipulator. *Int. J. Precis. Eng. and Manuf.* **17**, 11-17 (2016).

15. I. Cook, Newton's 'experimental' law of impacts. *The Math.Gazette* **70**, 107-114 (1986).

16. W. J. Stronge, *Impact Mechanics* (Cambridge U.P. Cambridge 2004)

17. R. Zwanzig, Rotational friction coefficients of a bumpy cylinder with slipping and sticking boundary conditions. *J. Chem. Phys.* **68**, 4325-4326 (1978).

18. H. P. Zhu, Z. Y. Zhou, R. Y. Yang and A. B. Yu, Discrete particle simulation of particulate systems: A review of major applications and findings. *Chem. Eng. Sci.* **63**, 5728-5779 (2008).

19. L. E. Sibert, D. Ertas, G. S. Grest, T. C Halsey, D. Levine and S. J. Plimpton, Granular flow down an inclined plane: Bagnold scaling and rheology. *Phys. Rev. E* **64**, 051302 (2001).

20. Plimpton, S. J. Fast parallel algorithms for short-range molecular dynamics. *J. Comp. Phys.* **117**, 1–19 (1995).

21. F. S. Crawford, Dynamics impedance matching with a lever, *Amer. J. Phys.* **57**, 52-56 (1989).

22. R.L.Garwin, Kinematics of an ultraelastic rough ball. *Amer. J. Phys.* **37**, 88-92 (1969)

23. S. Chapman and T. G. Cowling, *The mathematical theory of non-uniform gases.* (Cambridge UP, London, 1953), Chp. 11.

24. J. Odell and B. J. Berne, Molecular-dynamics of rough sphere fluid 1. Rotational relaxation. *J. Chem. Phys.* **63**, 2376-2394 (1975).

25. B. J. Berne and J. A. Montgomery, Coupling between translational and rotational motions, *Mol. Phys.* **32**, 363-378 (1976).

26. J. A. Montgomery and B. J. Berne, Viscoelastic theory of angular velocity correlation function. *J. Chem. Phys.* **66**, 2161-2165 (1977).

27. https://www.youtube.com/watch?v=0Vk7Qm87hUw

28. H. Brody, That's how the ball bounces, *Phys. Teach.* **22**, 494-497 (1984).